\documentclass[5p,sort&compress]{elsarticle}

\usepackage[utf8]{inputenc}
\usepackage{amsmath}
\usepackage{amssymb}
\usepackage[pdftex,colorlinks]{hyperref}
\usepackage{mathtools}
\usepackage{xcolor}

\DeclareMathOperator*{\Res}{Res}
\newcommand{\imag}{\text{i}}

\author[1]{Martin A.~Mojahed\corref{cor1}\fnref{fn1}}
\author[2]{Tom\'a\v{s} Brauner\fnref{fn2}}

\cortext[cor1]{Corresponding author}
\fntext[fn1]{Email: martinmoja96@gmail.com}
\fntext[fn2]{Email: tomas.brauner@uis.no}
\address[1]{Department of Physics, Norwegian University of Science and Technology, Hoegskoleringen 5, N-7491 Trondheim, Norway}
\address[2]{Department of Mathematics and Physics, University of Stavanger, N-4036 Stavanger, Norway}

\title{On-Shell Recursion Relations for Nonrelativistic Effective Field Theories}

\begin{document}

\begin{abstract}
We derive on-shell recursion relations for nonrelativistic effective field theories (EFTs) with enhanced soft limits. The recursion relations are illustrated through analytic calculation of tree-level scattering amplitudes in theories with a complex Schr\"odinger-type field, real scalar with linear dispersion relation, and real scalar with Lifshitz-type dispersion relation. Our results show that the landscape of gapless nonrelativistic EFTs with local $S$-matrix can be constrained by soft theorems and the consistency of the low-energy $S$-matrix similarly to massless relativistic EFTs. 
\end{abstract}

\maketitle


\section{Introduction}
\label{sec:intro}

On-shell recursion is a procedure to determine all scattering amplitudes in a theory recursively from a finite set of ``seed'' amplitudes. It plays a central part in the modern $S$-matrix program where physical and mathematical properties of scattering amplitudes are used to construct the $S$-matrix directly without the aid of a Lagrangian. Originally developed in the context of gauge theory by Britto, Cachazo, Feng and Witten (BCFW)~\cite{BCFW}, on-shell recursion was soon generalized to gravity theories~\cite{Gravity}, string theory~\cite{String}, generic renormalizable and some nonrenormalizable theories~\cite{(non)ren}. More recently, there has also been progress towards an on-shell formulation of scattering amplitudes in effective field theories (EFTs)~\cite{NLSMrecursion,CheungRecursion,LuoWen,Periodic}.

Beyond providing an efficient tool for calculating scattering amplitudes, recursion relations have also been successfully utilized as a framework to explore and classify the landscape of possible EFTs~\cite{Periodic,Elvang,SoftSmatrix,Shifman}. This connects to the newly emerging paradigm that seeks to \emph{define} quantum (effective) field theory without reference to a Lagrangian. While the basic principles underlying this program are mere locality and unitarity, the bulk of work done so far has focused on the sector of Lorentz-invariant field theories.\footnote{The only exceptions we are aware of include several recent works in a cosmological context, limited to EFTs with Lorentz-invariant kinematics but Lorentz-breaking interactions~\cite{boostlessbootstrap}, and a specific application of recursion techniques to scattering of phonons in Navier-Stokes fluids~\cite{CheungNavier}.} Yet, recent years have witnessed the EFT framework claiming a much larger territory than originally conceived. The range of novel applications of quantum field theory without Lorentz invariance now stretches from nonrelativistic gravity~\cite{Horava} and spacetime geometry~\cite{Son} to previously unthinkable exotic phases of quantum matter (see e.g.~Refs.~\cite{fractonGromov,fractonSeiberg} and references therein).

Should the modern scattering amplitude program provide new fundamental insight into the very nature of quantum field theory, it therefore seems mandatory to extend the scope of discussion by giving up on Lorentz invariance altogether. The aim of the present letter is to initiate the exploration of this new terra incognita. Our main result is that the existing on-shell recursion approach to EFT can be modified to nonrelativistic EFTs with rotationally-invariant gapless kinematics, where energy is proportional to an in principle arbitrary (integer) power of momentum. We demonstrate this by explicit examples of EFTs for a complex Schr\"odinger scalar and a real Lifshitz scalar.

The plan of the text is as follows. In the remainder of this section, we first briefly overview the BCFW recursion approach and its modification applicable to EFTs, and then outline the landscape of nonrelativistic EFTs relevant to our discussion. Sections~\ref{sec:deformation} and~\ref{sec:recursion} constitute the core of this letter, showing how to set up the recursion procedure for EFTs with nonrelativistic kinematics. An integral part of the text is section~\ref{sec:example} where we work out three examples.


\subsection{BCFW on-shell recursion}
\label{subsec:BCFWrecursion}

A central idea of the on-shell recursion technology is to promote $n$-particle on-shell amplitudes $A_n$ to meromorphic functions by complexifying external momenta in a way that preserves both on-shellness and conservation of energy and momentum. In the BCFW recursion, two selected external momenta, $p_i$ and $p_j$, are shifted,
\begin{align}
    \label{BCFWshift}
    \hat{p}_i\equiv p_i+zq, \qquad \hat{p}_j\equiv p_j-zq, \qquad z\in \mathbb{C}.
\end{align}
(Shifted quantities are denoted with a hat.) The auxiliary momentum $q$ must satisfy the on-shell conditions $q^2=p_i\cdot q=p_j\cdot q=0$. In four spacetime dimensions, it is thus fixed up to rescaling. At tree level, the complexified amplitude $\hat A_n(z)$ is a rational function of $z$. The original, physical amplitude $A_n=\hat A_n(0)$ can be recovered by

\begin{align}
    A_n=\frac{1}{2\pi\imag}\oint dz\frac{\hat{A}_n(z)}{z}, 
\end{align}
where the integration contour is an infinitesimal circle enclosing the origin of the complex plane. Cauchy's theorem and factorization then relate the physical amplitude $A_n$ to lower-point amplitudes in the following way, 
\begin{align}
    A_n&=-\sum_I\Res_{z=z_I}\frac{\hat{A}_n(z)}{z}+B_n\nonumber \\
    &=\sum_I\frac{\hat{A}_L^{(I)}(z_I)\hat{A}_R^{(I)}(z_I)}{P_I^2}+B_n.
\end{align}
The sum runs over all factorization channels $I$ where the lower-point amplitudes $\hat{A}_L^{(I)}$ and $\hat{A}_R^{(I)}$ contain one of $\hat p_i$, $\hat p_j$ each. Moreover, $P_I$ is the intermediate momentum evaluated at $z=0$, and $z_I$ is fixed by the on-shell condition $\hat P_I^2(z_I)=0$ to $z_I=-P_I^2/(2P_I\cdot q)$. Finally, $B_n$ denotes the contribution of the residue of the pole at $z=\infty$. The validity of the recursion relies on the latter either vanishing or being calculable.\footnote{Calculating $B_n$ is a challenging problem that has been considered in several contexts~\cite{Boundary}.} 

The above approach does not extend straightforwardly to low-energy EFTs. Technically, the problem is that the derivative couplings of EFTs imply polynomial growth of scattering amplitudes at large $z$, and thus preclude the standard recursion procedure. A different kind of complexification of the kinematical phase space is needed.


\subsection{On-shell recursion for EFTs}
\label{subsec:EFTrecursion}

The deeper reason why BCFW recursion fails for EFTs is that factorization alone is not sufficient to relate higher-point EFT amplitudes to lower-point ones; more information is needed. Since the form of an EFT is largely dictated by symmetries, it is hardly surprising that the additional input comes from symmetry (breaking).

Spontaneous symmetry breaking constrains the scattering amplitudes of the associated Nambu-Gold\-sto\-ne (NG) boson(s) in the ``(single) soft limit,'' in which the momentum of one of the particles participating in the scattering process vanishes. This limit can be probed by rescaling the momentum of the chosen particle, $p_i$, as $p_i\to \epsilon p_i$, and taking the scaling parameter $\epsilon$ to zero. The asymptotic behavior of the amplitude $A_n$ is characterized by a single scaling exponent,
\begin{align}
    \label{scaling}
    A_n\propto \epsilon^{\sigma_i}, \qquad \epsilon\to 0.
\end{align}
As a rule, albeit not without exceptions~\cite{Shifman}, spontaneous symmetry breaking ensures that $\sigma_i\geq1$; this fact is known as ``Adler's zero.'' Theories where $\sigma_i$ is larger than naively expected from counting derivatives in the Lagrangian are dubbed ``exceptional.'' The landscape of Lorentz-invariant exceptional EFTs is very strongly constrained~\cite{CheungEFT,Periodic,Bogers}. Single-flavor scalar exceptional EFTs were the first effective theories shown to be on-shell constructible~\cite{CheungRecursion} by a modification of the BCFW recursion procedure known as ``soft recursion.''

In the soft recursion procedure, \emph{all} external momenta are shifted,
\begin{align}
    \label{softrel}
    \hat{p}_i&\equiv p_i(1-a_iz), \qquad z\in \mathbb{C},\\
    \label{relcon}
    \sum_{i=1}^na_ip_i&=0,
\end{align}
where Eq.~(\ref{relcon}) is imposed by energy and momentum conservation. Nontrivial solutions for the coefficients $a_i$ exist for generic kinematical configurations when $n\geq D+2$, where $D$ is the spacetime dimension. The soft limit for the $i$-th particle can then be accessed by taking $z\to1/a_i$.

In order to be able to apply Cauchy's theorem, one modifies the behavior of the complexified amplitude $\hat A_n(z)$ at large $z$ by dividing it by the factor
\begin{align}
    \label{Fdef}
    F_n(z)\equiv\prod_{i=1}^n(1-a_iz)^{\sigma_i}.
\end{align}
For exceptional EFTs, this is sufficient to ensure vanishing of the boundary term $B_n$~\cite{CheungRecursion}. At the same time, the scaling~(\ref{scaling}) of the amplitude in the soft limit guarantees that adding $F_n(z)$ does not create any new poles in $\hat A_n(z)$. One can then reconstruct the physical amplitude $A_n=\hat A_n(0)$ similarly to the BCFW recursion,
\begin{align}
    \label{softAn}
    A_n=\frac{1}{2\pi\imag}\oint dz\frac{\hat{A}_n(z)}{zF_n(z)}=-\sum_I\Res_{z=z_I^{\pm}}\frac{\hat{A}_n(z)}{zF_n(z)},
\end{align}
where each factorization channel $I$ now gives rise to two poles $z_I^{\pm}$ corresponding to solutions of the shifted on-shell condition $\hat{P}^2_I(z)=0$. These are given explicitly by
\begin{align}
    \label{zpm}
    z_I^\pm=\frac1{Q_I^2}\Bigl[P_I\cdot Q_I\pm\sqrt{(P_I\cdot Q_I)^2-P_I^2Q_I^2}\Bigr],
\end{align}
where $P_I\equiv\sum\limits_{i\in I}p_i$ and $Q_I\equiv\sum\limits_{i\in I}a_ip_i$. Factorization together with Eq.~(\ref{softAn}) then imply the recursion formula~\cite{CheungRecursion}
\begin{align}
    \label{eq10}
    A_n=\sum_I\frac{\hat{A}_L^{(I)}(z_I^-)\hat{A}_R^{(I)}(z_I^-)}{P_I^2\left(1-\frac{z_I^-}{z_I^+}\right)F_n(z_I^-)}+(z_I^-\leftrightarrow z_I^+).
\end{align}


\subsection{Nonrelativistic EFTs}
\label{subsec:NREFTs}

The theories we will focus on in this letter live in a flat spacetime of $D\equiv d+1$ dimensions. They enjoy invariance under spacetime translations and $d$-dimensional spatial rotations. This is a fairly general setup that admits, if desired, a variety of kinematical algebras~\cite{Bacry}. The latter include the static (or Aristotelian) algebra containing no boosts whatsoever, and the Poincar\'e, Galilei (and its central extension, Bargmann) and Carroll algebras featuring different implementations of the relativity principle.

The NG modes stemming from spontaneous breakdown of global symmetry in such theories can be classified into two families, referred to as type $A_m$ and type $B_{2m}$ with positive integer $m$~\cite{HoravaShift}. A NG mode from the first family is described by a real scalar field with dispersion relation $\omega^2\propto\boldsymbol{p}^{2m}$. A NG mode from the second family, on the other hand, is described by two real scalar fields (or one complex scalar) forming a canonically conjugated pair with dispersion relation $\omega\propto \boldsymbol{p}^{2m}$.

Whether or not NG modes belonging to the $A_m$ and $B_{2m}$ families can exist in a given spatial dimension $d$ is constrained by the nonrelativistic version of the Cole\-man-Hohenberg-Mermin-Wagner (CHMW) theorem~\cite{HoravaNatural,WatanabePRX}. In short, at zero temperature, a NG boson of type $A_m$ may exist only if $m<d$. For fixed $m$, this in turn gives a lower bound on the dimension of space $d$. On the contrary, type $B_{2m}$ NG modes are not constrained at all and can exist, at zero temperature, for any positive $d$ and $m$.

It was observed early on~\cite{CheungEFT} that the enhanced scaling~\eqref{scaling} of scattering amplitudes in exceptional EFTs is a consequence of hidden symmetry. Motivated by this observation, one of us mapped in Ref.~\cite{NRLie} the landscape of nonrelativistic EFTs that admit such a hidden symmetry. We will show in a forthcoming paper~\cite{us} that unlike in the Lorentz-invariant case, this is in fact not sufficient to guarantee that a given EFT is exceptional. The catalogue of candidate EFTs compiled in Ref.~\cite{NRLie} will nevertheless serve as a useful guide for construction of explicit examples of nonrelativistic EFTs via recursion in section~\ref{sec:example}. We will thus be able to give examples of theories of the $A_1$, $A_2$ and $B_2$ type. Before doing so, we however first need to establish the soft recursion procedure for nonrelativistic EFTs. This is the subject of the next two sections.
      

\section{Momentum deformation in nonrelativistic EFTs}
\label{sec:deformation}

In this section, we introduce the momentum shifts nee\-ded for soft recursion. In contrary to the relativistic momentum shift in Eq.~(\ref{softrel}), we first shift the spatial momenta $\boldsymbol{p}_i$ only, and then use the on-shell condition to define an appropriate shift of the energies. 


\subsection{Soft shifts for type $B_{2m}$ theories}
\label{subsec:B2m}

The following shifts respect the on-shell condition for type $B_{2m}$ theories,  
\begin{align}
    \boldsymbol{\hat{p}}_i&\equiv \boldsymbol{p}_i(1-a_iz),\\
    \hat{p}^0_i&\equiv\boldsymbol{\hat{p}}_i^{2m}=\boldsymbol{p}_i^{2m}(1-a_iz)^{2m}.
\end{align}
Momentum and energy conservation then impose respectively the following constraints on the $a_i$ coefficients,
\begin{align}
    \label{B2m1}
    \sum_{i=1}^na_ie_i\boldsymbol{p}_i&=0, \\
    \label{B2m2}
    \sum_{i=1}^n(1-za_i)^{2m}e_i\boldsymbol{p}_i^{2m}&=0.
\end{align}
Here $e_i$ denotes a sign, chosen so that $e_i=+1$ for particles in the final state and $e_i=-1$ for particles in the initial state. Similarly to the relativistic case reviewed in section~\ref{subsec:EFTrecursion}, the existence of nontrivial solutions to Eq.~(\ref{B2m1}) requires $n\geq d+2$. Equation~(\ref{B2m2}) then imposes $2m$ additional constraints. Only amplitudes with $n\geq d+2+2m$ may therefore be reconstructed using soft recursion. For given $d$ and $m$, this tells us how many seed amplitudes we need to initiate the recursion procedure.


\subsection{Soft shifts for type $A_m$ theories}
\label{subsec:Am}

For type $A_{m}$ theories we define analogously
\begin{align}
    \boldsymbol{\hat{p}}_i&\equiv \boldsymbol{p}_i(1-a_iz),\\
    \hat{p}^0_i&\equiv |(\boldsymbol{p}_i^{2m})^{1/2}|(1-a_iz)^m,
\end{align}
which preserves on-shellness and yields the following constraints from momentum and energy conservation,
\begin{align}
    \label{ca1}
    \sum_{i=1}^na_ie_i\boldsymbol{p}_i&=0, \\
    \label{ca2}
    \sum_{i=1}^n (1-za_i)^me_i|(\boldsymbol{p}_i^{2m})^{1/2}|&=0.
\end{align}
Analogously to the type $B_{2m}$ case, the existence of nontrivial solutions for $a_i$ requires $n\geq d+2+m>2+2m$, where the last inequality follows from the nonrelativistic CHMW theorem. For the special case of $m=1$, which includes the family of Lorentz-invariant theories, the above constraints become equivalent to Eq.~(\ref{relcon}) and we recover the relativistic bound $n\geq d+3=D+2$.

Note that for both type $A_m$ and type $B_{2m}$ theories, the manifold of solutions for the $a_i$ coefficients is invariant under overall rescaling, $a_i\to\lambda a_i$, and overall shift, $a_i\to a_i+c$. This guarantees that in the special case of type $A_1$ theories where all the constraints on $a_i$ are linear, possible solutions for $a_i$ span an affine space.


\section{Soft recursion}
\label{sec:recursion}

We argued in section~\ref{subsec:EFTrecursion} that for relativistic exceptional EFTs, recursion relations among scattering amplitudes may be set up using Eq.~(\ref{softAn}). Since the argument only depends on the assumed soft behavior of $A_n$, factorization and vanishing of the boundary term, it can be generalized to any theory with these properties. Specifically, for theories of type $A_m$ and $B_{2m}$ we obtain
\begin{align}
    \label{NRrec}
    A_n=-\sum_I\sum_{i=1}^{2m}\Res_{z=z_I^{i}}\frac{\hat{A}_n(z)}{zF_n(z)}.
\end{align}
Here $z_I^i$, $i=1,\dotsc,$ $2m$ are solutions to the on-shell condition, which is of algebraic order $2m$ in $z$,
\begin{align}
    \bigl(\hat{P}_I^0\bigr)^2-\hat{\boldsymbol{P}}_I^{2m}&=0 && \text{for } A_m,  \\
    \hat{P}_I^0-\hat{\boldsymbol{P}}_I^{2m}&=0 && \text{for } B_{2m},
\end{align}
for a given factorization channel $I$, where compared to Eq.~(\ref{eq10}), $P_I$ is now defined with the appropriate signs $e_i$ where necessary. Factorization then implies that the amplitude~(\ref{NRrec}) can be expressed in terms of lower-point amplitudes,
\begin{align}
    \label{Angeneral}
    A_n=-\sum_I\sum_{i=1}^{2m}\Res_{z=z_I^{i}}\frac{\hat{A}_L^{(I)}(z)\hat{A}_R^{(I)}(z)}{zF_n(z)D^{(I)}(z)},
\end{align}
where
\begin{align}
    D^{(I)}(z)&=\bigl(\hat{P}_I^0\bigr)^2-\hat{\boldsymbol{P}}_I^{2m} && \text{for } A_m,  \\
    D^{(I)}(z)&=\hat{P}_I^0-\hat{\boldsymbol{P}}_I^{2m} && \text{for } B_{2m}.
\end{align}
Notice that the contribution from factorization channel $I$ in Eq.~(\ref{Angeneral}) matches the residue at $z=z_I^i$ of the following meromorphic function
\begin{align}
    \label{analyticf}
    \frac{\hat{A}_L^{(I)}(z)\hat{A}_R^{(I)}(z)}{zF_n(z)D^{(I)}(z)}.
\end{align}
This function can also have nonvanishing residues at $z=1/a_i$ and $z=0$. This follows from the fact that the intermediate propagator $D^{(I)}(z)$, hence also the subamplitudes $\hat{A}_L^{(I)}(z)$ and $\hat{A}_R^{(I)}(z)$, is off-shell for $z\neq z_I^i$. The on-shell argument implying that the soft behavior of the amplitudes dictated by Eq.~(\ref{scaling}) cancels the zeros of $F_n(z)$ is then no longer valid. In the special case where $\hat{A}_L^{(I)}(z)$ and $\hat{A}_R^{(I)}(z)$ are both local functions of momenta (that is, they have no poles) we can apply Cauchy's theorem to the meromorphic function in Eq.~(\ref{analyticf}) and recast the amplitude~(\ref{Angeneral}) in terms of a sum over residues at $z=0$ and $z=1/a_i$,
\begin{align}
    A_n={}&\sum_I\frac{\hat{A}_L^{(I)}(0)\hat{A}_R^{(I)}(0)}{D^{(I)}(0)}\nonumber\\
    &+\sum_I\sum_{i=1}^n\Res_{z=1/a_i}\frac{\hat{A}_L^{(I)}(z)\hat{A}_R^{(I)}(z)}{zF_n(z)D^{(I)}(z)}\nonumber \\
    \label{master}
    \equiv{}& A_n^{\text{ch}}+A_n^{\text{ct}}.
\end{align}
This expression is particularly useful for concrete applications. In terms of Feynman diagrams, the first term corresponds to the sum over diagrams with an internal propagator, whereas the second (double) sum encodes contributions from $n$-point contact operators. The two different types of contributions are distinguished by the notation introduced in the last line of Eq.~(\ref{master}).


\subsection{Validity criterion}
\label{subsec:validity}

Thus far we have simply assumed that the boundary term $B_n$ vanishes. A sufficient condition for this to happen is that $\hat{A}_n(z)/F_n(z)\to 0$ as $z\to\infty$. A criterion for the latter was in turn given by Elvang et al.~in  Ref.~\cite{Elvang}. Their argument only relies on dimensional analysis, the soft behavior of $A_n$, the analytic structure of tree-level amplitudes, and the freedom to shift all $a_i$ by an overall constant. Since the latter property survives in all type $A_m$ and $B_{2m}$ theories, as shown in section~\ref{sec:deformation}, it is easy to adapt the argument of Ref.~\cite{Elvang} for our purposes.

We start with a generic expression for the $n$-point tree-level amplitude,
\begin{align}
    A_n=\sum_j\Bigl(\prod_kg_k^{n_{jk}}\Bigr)M_j,
\end{align}
where $M_j$ are functions of momenta and $g_k$ are coupling constants associated with fundamental operators in the Lagrangian. Fundamental operators are defined in turn as the lowest-dimension operators whose on-shell matrix elements are needed to derive, at the leading-order in the low-energy expansion, any tree-level amplitude in the theory by recursion. Following the line of reasoning of Ref.~\cite{Elvang} then leads to the generalized validity criterion
\begin{align}
    \label{criterion}
    [A_n]-\min_j\Bigl(\sum_k n_{jk}[g_j]\Bigr)-\sum_{i=1}^n\sigma_i<0,
\end{align}
where square brackets indicate scaling dimension with respect to a uniform rescaling of all the momenta $\boldsymbol{p}_i$. It is easy to check that the criterion~(\ref{criterion}) is satisfied by all the example theories presented in the next section. 


\section{Example calculations}
\label{sec:example}

We will now work out three simple analytical examples of recursive reconstruction of scattering amplitudes in theories of type $B_2$, $A_1$ and $A_2$, respectively. All three sample theories feature tree-level amplitudes with soft scaling $\sigma_i=2$. Yet, each of the theories possesses Lagrangian representations with less than two derivatives per field, which means that they possess enhanced soft limits. We will show in a forthcoming paper~\cite{us} that the enhanced scaling of scattering amplitudes in these theories is a consequence of an interplay of spontaneously broken symmetry and dispersion relations of NG bosons. Each of the three theories contains just one physical NG mode. Since we no longer have to distinguish different $\sigma_i$ for different particles participating in the scattering process, we introduce a shorthand notation replacing Eq.~(\ref{Fdef}),
\begin{align}
    F_n^{(\sigma)}(z)\equiv \prod_{i=1}^n(1-a_iz)^{\sigma}.
\end{align}


\subsection{$B_2$: Schr\"odinger-DBI theory}
\label{subsec:SDBI}

Our first example features a complex scalar field $\Phi$ endowed with the action
\begin{align}
    \label{SDBI}
    S={}&\int\text{d}t\,\text{d}^d\boldsymbol{x}\, \bigl(\Phi^{\dagger}\imag\partial_0\Phi+\sqrt{G}-1\bigr), \\
    \label{Gmetric}
    G\equiv{}&1-2\boldsymbol\nabla\Phi\cdot\boldsymbol\nabla\Phi^{\dagger}+\bigl(\boldsymbol\nabla\Phi\cdot\boldsymbol\nabla\Phi^{\dagger}\bigr)^2\nonumber \\
    &-\bigl(\boldsymbol\nabla\Phi\cdot\boldsymbol\nabla\Phi\bigr)\bigl(\boldsymbol\nabla\Phi^{\dagger}\cdot\boldsymbol\nabla\Phi^{\dagger}\bigr).
\end{align}
This is a minimal nonrelativistic modification of one of the very few relativistic single-flavor exceptional theories~\cite{CheungEFT}: the Dirac-Born-Infeld (DBI) theory. We therefore name it the ``Schr\"odinger-DBI'' (SDBI) theory.

Our SDBI theory can be interpreted as describing fluctuations of a $d$-dimensional brane embedded in a $(d+2)$-dimensional Euclidean space. The symmetry of the SDBI action~(\ref{SDBI}) is accordingly $\mathbb{R}\times\text{ISO}(d+2)$, with the first factor of $\mathbb{R}$ corresponding to time translations~\cite{NRLie}. This symmetry is spontaneously broken down to $\mathbb{R}\times\text{ISO}(d)\times\text{SO}(2)$ by the presence of the brane, and the real and imaginary parts of $\Phi$ correspond to NG fields of spontaneously broken translations in the two extra dimensions. The term in Eq.~(\ref{SDBI}) with a single time derivative is only invariant under the full symmetry up to a surface term. It is thus an example of a Wess-Zumino-Witten (WZW) term.

The action~(\ref{SDBI}) fixes all tree-level amplitudes. We will now demonstrate that the recursion formula~(\ref{master}) correctly reproduces the six-point amplitude starting from the seed four-point amplitude. In fact, the argument of section~\ref{subsec:B2m} limits the validity of the recursion for $n=6$ to $d\leq2$ spatial dimensions. However, the amplitudes $A_n$ as functions of the momenta $\boldsymbol{p}_i$ do not depend explicitly on $d$. Whatever analytic relations between the amplitudes we find will therefore be independent of $d$ as well. One may think of this as carrying out the recursive step from $A_4$ to $A_6$ in $d=2$ dimensions, and then analytically continuing the result to any value of $d$ of interest.

To make the calculation transparent, we first explicitly list the relevant parts of the Lagrangian,
\begin{align}
    &\mathcal{L}_2=\Phi^{\dagger}(\imag\partial_0+\boldsymbol\nabla^2)\Phi, \\
    \label{SL4}
    &\mathcal{L}_4=-\frac{1}{2}\bigl(\boldsymbol\nabla\Phi\cdot\boldsymbol\nabla\Phi\bigr)\bigl(\boldsymbol\nabla\Phi^{\dagger}\cdot\boldsymbol\nabla\Phi^{\dagger}\bigr), \\
    \label{L6}
    &\mathcal{L}_6=-\frac{1}{2}\bigl(\boldsymbol\nabla\Phi\cdot\boldsymbol\nabla\Phi\bigr)\bigl(\boldsymbol\nabla\Phi\cdot\boldsymbol\nabla\Phi^{\dagger}\bigr)\bigl(\boldsymbol\nabla\Phi^{\dagger}\cdot\boldsymbol\nabla\Phi^{\dagger}\bigr).
\end{align}
Charge conservation dictates that the numbers of incoming and outgoing Schr\"odinger scalars must match in any scattering process. We use the convention that the particles labeled $1,\dotsc,n/2$ are incoming, whereas the particles $n/2,\dotsc,n$ are outgoing. The seed on-shell four-point amplitude then follows immediately from Eq.~(\ref{SL4}) as
\begin{align}
    A_4=2(\boldsymbol{p}_1\cdot\boldsymbol{p}_2)(\boldsymbol{p}_3\cdot \boldsymbol{p}_4).
\end{align}

We are now ready to derive the six-point amplitude by recursion. We will use the indices $a$, $b$, $c$ to label a permutation of the incoming particles and $d$, $e$, $f$ a permutation of the outgoing particles such that $a$, $b$, $f$ are on the same side of the factorization channel. We can then identify the nine factorization channels in terms of $c$ and $f$ alone, 
\begin{align}
    \nonumber
    I=\{(c,f)\}=\{&(14),(15), (16), (24),(25), (26),\\
    &(34), (35), (36) \}.
\end{align}
Energy and momentum conservation fix the parameters of the intermediate propagator for each factorization channel,
\begin{align}
    \boldsymbol{P}_I&\equiv\boldsymbol{p}_a+\boldsymbol{p}_b-\boldsymbol{p}_f=\boldsymbol{p}_d+\boldsymbol{p}_e-\boldsymbol{p}_c, \\
    \frac{1}{2}\left(P_I^0-\boldsymbol{P}_I^2\right)&=-\boldsymbol{p}_a\cdot\boldsymbol{p}_b-\boldsymbol{p}_f\cdot\boldsymbol{p}_f+\boldsymbol{p}_a\cdot\boldsymbol{p}_f+\boldsymbol{p}_b\cdot\boldsymbol{p}_f\nonumber \\
    &=-\boldsymbol{p}_d\cdot\boldsymbol{p}_e-\boldsymbol{p}_c\cdot\boldsymbol{p}_c+\boldsymbol{p}_d\cdot\boldsymbol{p}_c+\boldsymbol{p}_e\cdot\boldsymbol{p}_c.
    \nonumber
\end{align}
The channel contribution $A_6^{\text{ch}}$ as defined by Eq.~(\ref{master}) reads
\begin{align}
    \label{channeldbi}
    &A_6^{\text{ch}}=4\sum_I\frac{(\boldsymbol{p}_a\cdot\boldsymbol{p}_b)(\boldsymbol{p}_d\cdot\boldsymbol{p}_e)(\boldsymbol{p}_c\cdot\boldsymbol{P}_I)(\boldsymbol{p}_f\cdot\boldsymbol{P}_I)}{P_I^0-\boldsymbol{P}_I^2}\\
    &=\smashoperator{\sum_{\sigma,\rho\in S_3}}\frac{(\boldsymbol{p}_{\sigma(1)}\cdot\boldsymbol{p}_{\sigma(2)})(\boldsymbol{p}_{\rho(4)}\cdot\boldsymbol{p}_{\rho(5)})(\boldsymbol{p}_{\sigma(3)}\cdot\boldsymbol{k}_{\sigma\rho})(\boldsymbol{p}_{\rho(6)}\cdot\boldsymbol{k}_{\sigma\rho})}{k_{\sigma\rho}^0-\boldsymbol{k}_{\sigma\rho}^2},\nonumber
\end{align}
where $\sigma$ and $\rho$ denote respectively permutations of $\{1,2,3\}$ and $\{4,5,6\}$, and we have used the shorthand notation
\begin{align}
    \boldsymbol{k}_{\sigma\rho}\equiv \boldsymbol{p}_{\sigma(1)}+\boldsymbol{p}_{\sigma(2)}-\boldsymbol{p}_{\rho(6)}.
\end{align}
The second line of Eq.~(\ref{channeldbi}) is manifestly equal to the Feynman diagram expression one obtains from Eq.~(\ref{SL4}).

Similarly, the contact contribution to the six-point amplitude follows from Eq.~(\ref{master}) as
\begin{align}
    \label{contact1}
    A_6^{\text{ct}}&=4\sum_I\sum_{i=1}^6\Res_{z=z_i}\frac{(\boldsymbol{\hat{p}}_a\cdot\boldsymbol{\hat{p}}_b)(\boldsymbol{\hat{p}}_d\cdot\boldsymbol{\hat{p}}_e)(\boldsymbol{\hat{p}}_c\cdot\boldsymbol{\hat{P}}_I)(\boldsymbol{\hat{p}}_f\cdot\boldsymbol{\hat{P}}_I)}{zF_6^{(2)}(z)\bigl(\hat{P}_I^0-\boldsymbol{\hat{P}}_I^2\bigr)} \nonumber \\
    &\equiv -2\sum_I\sum_{i=1}^6f(z_i).
\end{align}
The residues at $z_i\equiv1/a_i$ for a given factorization channel can be rewritten as
\begin{align}
    f(z_a)&=\Res_{z=z_a}\frac{(\boldsymbol{\hat{p}}_a\cdot\boldsymbol{\hat{p}}_b)(\boldsymbol{\hat{p}}_d\cdot\boldsymbol{\hat{p}}_e)(\boldsymbol{\hat{p}}_c\cdot\boldsymbol{\hat{p}}_f-\boldsymbol{\hat{p}}_c\cdot\boldsymbol{\hat{p}}_b)}{zF_6^{(2)}(z)}, \nonumber\\
    f(z_b)&=\Res_{z=z_b}\frac{(\boldsymbol{\hat{p}}_a\cdot\boldsymbol{\hat{p}}_b)(\boldsymbol{\hat{p}}_d\cdot\boldsymbol{\hat{p}}_e)(\boldsymbol{\hat{p}}_c\cdot\boldsymbol{\hat{p}}_f-\boldsymbol{\hat{p}}_c\cdot\boldsymbol{\hat{p}}_a)}{zF_6^{(2)}(z)}, \nonumber\\
    f(z_c)&=\Res_{z=z_c}\frac{(\boldsymbol{\hat{p}}_a\cdot\boldsymbol{\hat{p}}_b)(\boldsymbol{\hat{p}}_c\cdot\boldsymbol{\hat{p}}_d+\boldsymbol{\hat{p}}_c\cdot\boldsymbol{\hat{p}}_e)(\boldsymbol{\hat{p}}_d\cdot\boldsymbol{\hat{p}}_f+\boldsymbol{\hat{p}}_e\cdot\boldsymbol{\hat{p}}_f)}{zF_6^{(2)}(z)}, \nonumber\\
    f(z_d)&=\Res_{z=z_d}\frac{(\boldsymbol{\hat{p}}_a\cdot\boldsymbol{\hat{p}}_b)(\boldsymbol{\hat{p}}_d\cdot\boldsymbol{\hat{p}}_e)(\boldsymbol{\hat{p}}_c\cdot\boldsymbol{\hat{p}}_f-\boldsymbol{\hat{p}}_e\cdot\boldsymbol{\hat{p}}_f)}{zF_6^{(2)}(z)}, \nonumber\\
   f(z_e)&=\Res_{z=z_e}\frac{(\boldsymbol{\hat{p}}_a\cdot\boldsymbol{\hat{p}}_b)(\boldsymbol{\hat{p}}_d\cdot\boldsymbol{\hat{p}}_e)(\boldsymbol{\hat{p}}_c\cdot\boldsymbol{\hat{p}}_f-\boldsymbol{\hat{p}}_d\cdot\boldsymbol{\hat{p}}_f)}{zF_6^{(2)}(z)}, \nonumber\\
    f(z_f)&=\Res_{z=z_f}\frac{(\boldsymbol{\hat{p}}_d\cdot\boldsymbol{\hat{p}}_e)(\boldsymbol{\hat{p}}_a\cdot\boldsymbol{\hat{p}}_c+\boldsymbol{\hat{p}}_b\cdot\boldsymbol{\hat{p}}_c)(\boldsymbol{\hat{p}}_a\cdot\boldsymbol{\hat{p}}_f+\boldsymbol{\hat{p}}_b\cdot\boldsymbol{\hat{p}}_f)}{zF_6^{(2)}(z)}.\nonumber
\end{align}
After substituting the expressions above into Eq.~(\ref{contact1}), collecting the contributions to the residue at each $z_i$ from all factorization channels, and using (shifted) momentum conservation, we obtain
\begin{align}
    A_6^{\text{ct}}=&-\frac{1}{2}\sum_{i=1}^6\Res_{z=z_i}\frac1{zF_6^{(2)}(z)}\\
    &\times\smashoperator{\sum_{\sigma,\rho\in S_3}}(\boldsymbol{\hat p}_{\sigma(1)}\cdot\boldsymbol{\hat p}_{\sigma(2)})(\boldsymbol{\hat p}_{\sigma(3)}\cdot\boldsymbol{\hat p}_{\rho(4)})(\boldsymbol{\hat p}_{\rho(5)}\cdot\boldsymbol{\hat p}_{\rho(6)}).\nonumber
\end{align}
A final application of Cauchy's theorem yields
\begin{align}
    A_6^{\text{ct}}=\frac{1}{2}\smashoperator{\sum_{\sigma,\rho\in S_3}}(\boldsymbol{p}_{\sigma(1)}\cdot\boldsymbol{p}_{\sigma(2)})(\boldsymbol{p}_{\sigma(3)}\cdot\boldsymbol{p}_{\rho(4)})(\boldsymbol{p}_{\rho(5)}\cdot\boldsymbol{p}_{\rho(6)}),
\end{align}
which is manifestly equal to the contribution from the contact term in Eq.~(\ref{L6}). 


\subsection{$A_1$: spatial Galileon}
\label{subsec:galileon}

Our second example includes a whole class of La\-gran\-gians of a real scalar field $\phi$, 
\begin{align}
    \label{spatialGalileon}
    \mathcal{L}=\frac12(\partial_\mu\phi)^2+\sum_{n=3}^{d+1}c_n\phi G_{n-1},
\end{align}
where $c_n$ are real coupling constants and $G_n$ is a polynomial of order $n$ in the second spatial derivatives of $\phi$,
\begin{align}
    G_n\equiv{}&\frac1{(d-n)!}\epsilon^{i_1\dotsb i_nk_{n+1}\dotsb k_d}\epsilon^{j_1\dotsb j_n}_{\phantom{j_1\dotsb j_n}k_{n+1}\dotsb k_d}\nonumber\\
    &\times(\partial_{i_1}\partial_{j_1}\phi)\dotsb(\partial_{i_n}\partial_{j_n}\phi).
\end{align}
This is a nonrelativistic version of another type of a relativistic single-flavor exceptional theory~\cite{CheungEFT}: the Galileon. As opposed to the usual, Lorentz-invariant Galileon theory~\cite{Gal}, the interaction part of Eq.~(\ref{spatialGalileon}) contains only spatial derivatives of $\phi$. We therefore dub it ``spatial Galileon.'' The action~(\ref{spatialGalileon}) is invariant under polynomial shifts of $\phi$ of first order in spatial coordinates, $\phi\to\phi+\alpha+\boldsymbol\beta\cdot\boldsymbol x$. This spatial version of the usual Galileon symmetry is a special case of a class of ``multipole algebras'' that have recently attracted attention in the context of fracton physics~\cite{fractonGromov}. All interaction terms in Eq.~(\ref{spatialGalileon}) as well as the spatial part of the kinetic term are of the WZW type~\cite{Goon}.

Since the spatial Galileon is a type $A_1$ theory, the validity of the recursion is limited to $n$-point amplitudes with $n\geq d+3$, as shown in section~\ref{subsec:Am}. For illustration, we will now restrict Eq.~(\ref{spatialGalileon}) to the quartic interaction term and show how to reconstruct the six-point amplitude. This requires setting $d=3$, since for $d<3$ the quartic spatial Galileon interaction does not exist.

It is convenient to express the Feynman rule for the $n$-point spatial Galileon vertex as~\cite{GalDual}
\begin{align}
    \label{VnGal}
    V_n(\boldsymbol p_1,\dotsc,\boldsymbol p_n)=c_n'\smashoperator{\sum_{\sigma\in Z_n}}G(\boldsymbol p_{\sigma(1)},\dotsc,\boldsymbol p_{\sigma(n-1)}),
\end{align}
where $G(\boldsymbol p_1,\dotsc,\boldsymbol p_{n-1})$ is the Gram determinant, that is the determinant of the $(n-1)\times(n-1)$ matrix with entries $\boldsymbol p_i\cdot\boldsymbol p_j$. Importantly, the Gram determinant is a symmetric, homogeneous polynomial of order two in all its arguments,
\begin{align}
    \label{gscaling}
    G(\lambda\boldsymbol p_1,\dotsc,\boldsymbol p_{n-1})=\lambda^2G(\boldsymbol p_1,\dotsc,\boldsymbol p_{n-1}).
\end{align}
Due to momentum conservation in the vertex, all the contributions to the sum in Eq.~(\ref{VnGal}) are then equal and we can write $V_n=nc_n'G(\boldsymbol p_1,\dotsc,\boldsymbol p_{n-1})$.

The six-point amplitude is now determined in terms of the four-point seed amplitude by Eq.~(\ref{master}),
\begin{align}
    \label{galA6}
    A_6=\sum_I\Biggl\{&\frac{A_{4L}^{(I)}A_{4R}^{(I)}}{(P_I^0)^2-\boldsymbol{P}_I^2}\nonumber\\
    &+\sum_{i=1}^6\Res_{z=1/a_i}\frac{\hat{A}_{4L}^{(I)}(z)\hat{A}_{4R}^{(I)}(z)}{zF^{(2)}_6(z)\bigl[(\hat{P}_I^0)^2-\boldsymbol{\hat{P}}^2_I\bigr]}\Biggr\}.
\end{align}
For a generic permutation $\sigma$ of the external momenta, the numerator in the last term can be cast as
\begin{align}
    &V_4(\boldsymbol{\hat p}_{\sigma(1)},\boldsymbol{\hat p}_{\sigma(2)},\boldsymbol{\hat p}_{\sigma(3)},\boldsymbol {\hat P}_I)V_4(\boldsymbol{\hat p}_{\sigma(4)},\boldsymbol{\hat p}_{\sigma(5)},\boldsymbol{\hat p}_{\sigma(6)},\boldsymbol {\hat P}_I)\nonumber\\
    &=(4c_4')^2G(\boldsymbol{\hat p}_{\sigma(1)},\boldsymbol{\hat p}_{\sigma(2)},\boldsymbol{\hat p}_{\sigma(3)})G(\boldsymbol{\hat p}_{\sigma(4)},\boldsymbol{\hat p}_{\sigma(5)},\boldsymbol{\hat p}_{\sigma(6)}).
\end{align}
The scaling property~(\ref{gscaling}) of the Gram determinant then ensures that the denominator factor $F^{(2)}_6(z)$ in Eq.~(\ref{galA6}) is canceled. Thus, all the residues inside the second sum in Eq.~(\ref{galA6}) vanish and only the first, ``channel'' term therein survives. This is manifestly equal to the expression for $A_6$ one obtains using Feynman diagrams.


\subsection{$A_2$: Lifshitz scalar with polynomial shift symmetry}
\label{subsec:lifshitz}

Our final example is a so-called $z=2$ Lifshitz theory, which possesses the following kinetic term,
\begin{align}
    \mathcal{L}_2=\frac{1}{2}\left(\partial_0\phi\right)^2-\frac{1}{2}\left(\boldsymbol\nabla^2\phi\right)^2.
\end{align}
This Lagrangian is strictly invariant under the spatial Ga\-li\-leon symmetry.\footnote{Lifshitz scalars with polynomial shift symmetries have been classified in Refs.~\cite{HinteShift,HoravaShift} and shown to exhibit rich and surprising features that shed new light on the concept of naturalness in nonrelativistic quantum field theory~\cite{HoravaShift,HoravaNatural}.} We can thus add the spatial Ga\-li\-leon interactions in Eq.~(\ref{spatialGalileon}) to it. The ensuing theory can be viewed as a fine-tuned version of the spatial Galileon where the usual kinetic term proportional to $(\boldsymbol\nabla\phi)^2$ is set to zero.

This is a type $A_2$ theory, so the validity of the recursion is limited to $n$-point amplitudes with $n\geq d+4$ as shown in section~\ref{subsec:Am}. At the same time, the CHMW theorem requires that $d>2$. We thus cannot reconstruct the six-point amplitude by recursion. We can however consider a seed five-point vertex and use recursion to reconstruct the eight-point amplitude. This requires setting $d=4$, since for $d<4$ the quintic spatial Galileon does not exist.

Following the same steps as in the previous example, Eq.~(\ref{master}) then gives the following result for the eight-point amplitude,
\begin{align}
    A_8=\sum_I\frac{A_{5L}^{(I)}A_{5R}^{(I)}}{(P_I^0)^2-\boldsymbol{P}_I^4},
\end{align}
which agrees with the Feynman diagram expression. 


\section{Outlook}
\label{sec:outlook}

We have derived recursion relations for nonrelativistic EFTs with enhanced soft limits. To the best of our knowledge, this is the first time that on-shell constructibility for theories without Lorentz invariance has been shown.

Beyond providing a new tool for calculating explicit tree-level amplitudes in specific field theories, soft recursion is a key ingredient in the ``soft bootstrap'' program, which explores and classifies the space of possible EFTs. In a paper soon to appear~\cite{us}, we will carry out a more detailed classification of possible seed amplitudes. When combined with soft recursion, this will allow us to perform a scan of the landscape of nonrelativistic EFTs, improving on our previous symmetry-based study~\cite{NRLie}.

Our recursion relations can also be applied to theories with universal albeit not necessarily vanishing soft behavior by following the line of reasoning in Ref.~\cite{LuoWen}. This would require new soft theorems for NG boson amplitudes~\cite{Shifman}, an avenue we leave open for future work.


\section*{Acknowledgements}
T.B.~would like to thank Andreas Helset for a discussion on a related subject. M.A.M.~acknowledges the hospitality of the University of Stavanger, where the majority of the work was done. This work has been supported by the grant no.~PR-10614 within the ToppForsk-UiS program of the University of Stavanger and the University Fund.


\end{document}